\documentclass[anonymous]{iopconfser}

\usepackage{quotes}
\usepackage{graphicx}
\usepackage{amsmath}  
\usepackage{amsfonts} 
\usepackage{xcolor}

\begin{document}

\title{Enhancing Kinematics Understanding Through a Real-Time Graph-Based Motion Video Game}

\author{Mateo Dutra$^{0000-0002-0251-8052}$, Marcos Abreu$^{0009-0000-6518-7695}$, Martín Monteiro$^{0000-0001-9472-2116, 3}$, Silvia Sguilla, Cecilia Stari$^{0000-0002-4349-1036, 1}$, Alvaro Suarez$^{0000-0002-5345-5565}$,  Arturo C. Mart{\'i}$^{0000-0003-2023-8676, 1}$ }

\affil{$^1$ Instituto de F\'{i}sica, Universidad de la
  Rep\'{u}blica, Montevideo, Uruguay}
\affil{$^2$ Consejo de Formacion en Educación, Montevideo, Uruguay}
\affil{$^3$ Universidad ORT Uruguay, Montevideo, Uruguay}

\email{marti@fisica.edu.uy}

\begin{abstract}
Kinematics is a core topic in early physics courses, yet students often struggle to interpret motion and its graphical representations. To tackle these difficulties, we developed MissionMotion, a physical–computational videogame where students reproduce target motion graphs using real-time data from their own movements or from sensors connected through micro:bit or Arduino. The system displays both the target and the user-generated graph, providing immediate visual feedback and a score based on similarity. We piloted the environment with ninth-grade students in different school contexts and evaluated their experience using the MEEGA+ instrument. The results show strong engagement, positive perceptions of usability, and evidence that the game promotes reflection on motion graphs in ways that rarely emerge in traditional lessons. MissionMotion runs on any web-enabled device and all materials are openly available, offering teachers an accessible tool to integrate experimentation, computational thinking, and playful learning into physics classrooms.
\end{abstract}
\section{A Broader View on Motion Learning in the Classroom}

The study of motion and its causes is fundamental to understanding how we interact with the world—from the operation of vehicles and the flight of a ball to human movement and the dynamics of the solar system. This central role is reflected in physics curricula across educational levels. Yet research consistently shows that many students struggle to grasp the fundamental concepts of kinematics even after years of instruction \cite{Ivanjek2016,mcdermott1987student,trowbridge1980investigation,trowbridge1981investigation}. In an educational landscape that increasingly values interdisciplinarity and meaningful engagement, teachers face the challenge of adopting approaches that promote deeper conceptual understanding while connecting physics to students’ everyday experiences.

In this project, we designed and implemented a physical–computational videogame environment supported by a set of activities and open resources. The approach is inspired by large-scale educational technology initiatives such as Uruguay’s Plan Ceibal, a nationwide program that provided every primary and secondary student with a laptop and free internet access, coupled with extensive teacher training and a robust national monitoring framework (for additional information see \texttt{http://ceibal.edu.uy}). Our system integrates programmable boards, sensors, and robotics kits to teach kinematics while fostering both scientific inquiry and computational thinking skills. Concretely, the main outcome is MissionMotion, a cross-platform web application that challenges students to reproduce motion graphs in real time: they control a paper-ball avatar moving along a ruler-like axis while its position or velocity is dynamically plotted next to a target graph. Unlike traditional simulations, the environment incorporates real-world sensor data and invites students to interact using their own body movements, mirroring actions suggested by the program under teacher guidance; depending on available resources, motion can be controlled via mouse, touchscreen, smartphone sensors, or external devices such as ultrasonic sensors connected through micro:bit or Arduino boards. Accessible through any web browser and compatible with mobile devices, tablets, and computers, the platform allows students to record their progress and share it with instructors.

The project is ongoing, and the first rounds of systematic evaluation indicate very positive outcomes in terms of engagement, usability, and students’ reflection on motion graphs. Evidence regarding conceptual learning gains in scientific and computational reasoning is currently being collected through standardized pre- and post-assessments and will be reported in future work. All resources are freely available as open educational materials in the website \texttt{http://missionmotion.uy}. In this paper, we expand upon the results presented at the GIREP 2025 Conference (GIREP–EPEC) held in Leiden, The Netherlands. Within this broader agenda, the present article focuses on the design of MissionMotion and on a pilot implementation with ninth-grade students in different school contexts, using the MEEGA+ instrument to analyze player experience, usability, and perceived educational value as a tool for working with motion graphs \cite{petri2016meega}.

\section{Theoretical Framework}

Motion provides a unifying framework for describing and explaining physical phenomena across a wide range of scales, from everyday human activities to complex technological and natural systems. Concepts such as position, velocity, and acceleration are essential for modeling how objects evolve in time and for establishing the foundations of classical mechanics. For this reason, kinematics plays a structuring role in physics education, particularly in secondary school curricula, where it is introduced as a central component of general physics courses, and later in university-level introductory physics, where it supports the development of more advanced ideas in mechanics and engineering.

Despite its importance and repeated inclusion in curricula, numerous studies have shown that students experience persistent difficulties in grasping basic kinematic concepts. These include confusions between position, displacement, distance, velocity, and acceleration~\cite{trowbridge1980investigation,trowbridge1981investigation,mcdermott1987student}. When interpreting graphs of position, velocity, or acceleration as functions of time, students often misinterpret graphs as literal pictures of motion, or mistake the slope at a point for the value on the vertical axis~\cite{beichner1994testing,Ivanjek2016,Duijzer2019}. Such conceptual challenges highlight the limitations of traditional instruction and underscore the need for new teaching strategies that promote active engagement and deeper conceptual understanding.

Active learning methodologies have consistently demonstrated substantial improvements in students’ conceptual understanding of physics and reasoning skills. In particular, real-time graphical visualization of motion has proven effective in helping students connect mathematical representations with physical experiences~\cite{Hake1998,aprendizajeactivoB,aprendizajeactivoC}. Another promising approach is \emph{gamification}, the use of game design elements in non-game contexts, which can enhance motivation and engagement in science education~\cite{aprendizajeactivoD}. By integrating these approaches, educators can help students not only learn abstract concepts but also experience and manipulate them through interactive, embodied activities.

In the Uruguayan educational context, \emph{Plan Ceibal} has played a transformative role in introducing digital tools and programmable hardware—such as micro:bit boards, tablets, 3D printers, and robotics kits—into schools. These technologies provide fertile ground for designing interactive and experimental activities in science education \cite{jara2018policies}. Furthermore, several studies have demonstrated that computational tools can support students’ understanding of physical phenomena by enabling data visualization and problem-solving in new ways \cite{hutchins2018studying}. As physics becomes increasingly computational in nature, introductory courses represent an ideal context for developing \emph{computational thinking} (CT) skills \cite{caballero2018prevalence}. Physics education standards worldwide are encouraging instructors to integrate computation as a fundamental component of scientific practice.

Computational thinking—defined as the ability to model real-world problems and design algorithmic solutions that can be computationally implemented—has emerged as a core 21st-century competency
\cite{wing2017computational}. Ceibal’s national initiatives have embraced this perspective, promoting CT as a transversal competence across subjects. Through programming and robotics, students engage with logical reasoning, creativity, collaboration, and problem-solving, applying these skills in diverse domains such as mathematics, language, and the sciences \cite{chevalier2020fostering}. This integration encourages interdisciplinary approaches where computational thinking becomes a driving force for developing conceptual understanding and procedural fluency in multiple areas of the curriculum.

Physics, as an experimental and model-based science, provides an ideal context for connecting computational thinking with scientific inquiry. Previous research has shown that integrating CT into physics instruction allows students to construct and test models, analyze data, and reflect on the relationships between physical quantities and computational representations \cite{orban2020computational,weller2022development,gambrell2024analyzing,dutra2025code}. These studies suggest that CT-oriented learning environments can foster both conceptual understanding and critical reasoning. However, questions remain regarding the most effective ways to embed these practices in school settings, particularly in lower and upper secondary levels where abstract reasoning skills are still developing \cite{basu2016identifying}.

Against this backdrop, our project explores whether \emph{gamification can serve as an effective pathway for integrating computational thinking into physics education}. Specifically, we developed a physical–computational videogame environment aimed at helping students overcome conceptual difficulties in kinematics and graphical interpretation while promoting engagement and motivation. The environment links real-world motion—captured through sensors—to a digital game interface, enabling students to visualize their own movements as graphs of position, velocity, or acceleration in real time. By comparing their motion-generated graphs with target graphs provided by the game, students receive immediate feedback and a performance score based on accuracy.

This dual physical–digital interaction bridges abstract representations with embodied experience, reinforcing conceptual understanding through active participation. The environment also supports multiple modes of use: a simple mode in which students control a virtual character using a mouse or keyboard, and more advanced modes that employ motion sensors or smartphone accelerometers to track real physical movement. Teachers may further extend the activity by having students program the sensors or robots that interface with the game, deepening their computational engagement and fostering problem-solving, debugging, and iterative design skills.

At the project level, the broader evaluation plan combines different sources of evidence. To assess learning outcomes, the project employs rubrics aligned with national science and computer science curricula, focusing on students’ abilities to interpret graphical data, relate physical quantities, apply logical and creative thinking, and program devices that interact with their environment. Additionally, pre- and post-assessment tools commonly used in physics education research are applied to evaluate students’ conceptual gains in kinematics. Together, these instruments define an evaluation agenda that extends beyond the scope of this article, which concentrates on an initial pilot implementation and on players’ experience with the environment.

To sump up, this theoretical framework situates the proposed project at the intersection of physics education research, gamification, and computational thinking. It responds to well-documented learning difficulties in kinematics by leveraging active learning, real-time visualization, and interactive digital technologies. The approach aligns with current educational reforms emphasizing interdisciplinary competencies and seeks to contribute evidence on how playful, technology-rich environments can foster both conceptual and computational understanding in physics.

\section{Description and activities}

\begin{figure}[ht]
 \centering
\includegraphics[width=0.85\textwidth]{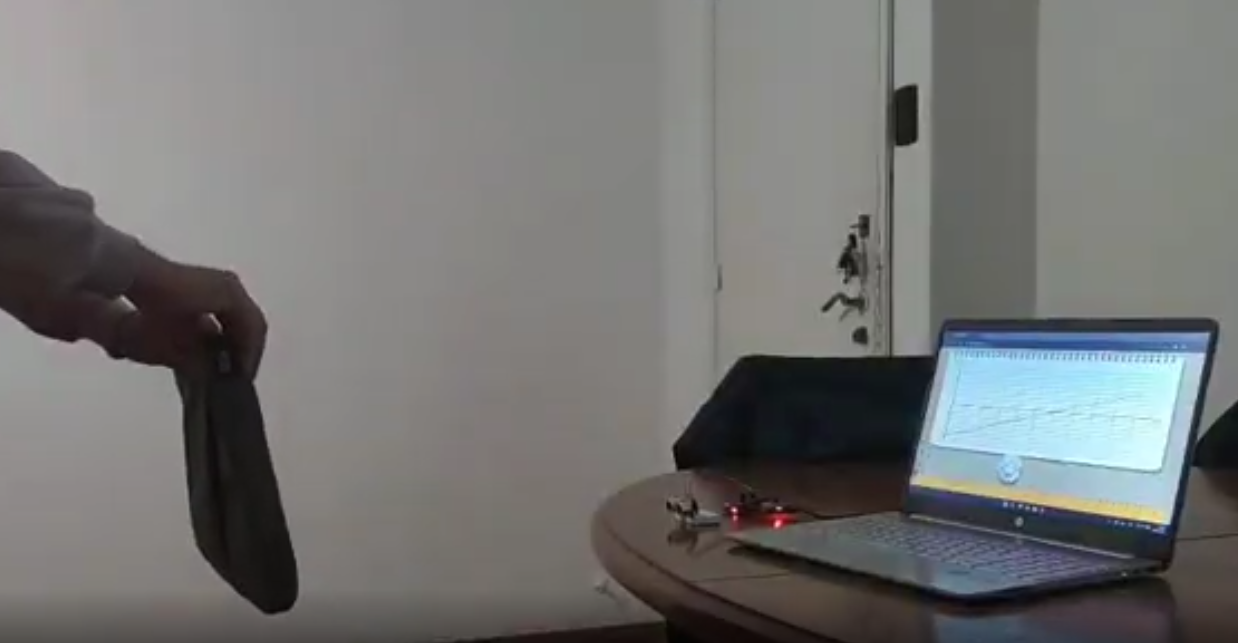}
 \caption{Micro:bit board linked to an ultrasonic sensor and interfaced with a laptop for real-time motion acquisition in MissionMotion. The photo also shows the player holding a folder used as a paddle to improve the ultrasonic sensor’s tracking accuracy.  \label{fig:micro}}
\end{figure}

The main project's outcome is MissionMotion, a cross-platform web application designed to improve conceptual understanding of kinematics by allowing students to reproduce motion graphs in real time. It operates as a physical–computational learning environment that bridges virtual interaction with real-world motion: students control a paper-ball avatar moving along a ruler-like axis while its position or velocity is dynamically plotted alongside a target graph they must imitate. Motion can be controlled via mouse, trackpad, touchscreen, smartphone sensors, or external ultrasonic sensors connected through micro:bit or Arduino, enabling seamless integration with existing Plan Ceibal kits in Uruguay. A scoring algorithm evaluates the similarity between the user-generated and reference graphs, providing immediate feedback. Beyond gameplay, the platform functions as an open-source resource hub offering sequenced activities, opportunities to program sensors or robots to foster computational thinking, publishable materials, and classroom implementation to systematically measure its educational impact. Its browser-based implementation ensures smooth operation across a wide range of devices without requiring any platform-specific installation or configuration.  An example of gameplay using a micro:bit board and an ultrasonic sensor is shown in Fig.~\ref{fig:micro}.

\begin{figure}[ht]
 \centering
\includegraphics[width=0.4\textwidth]{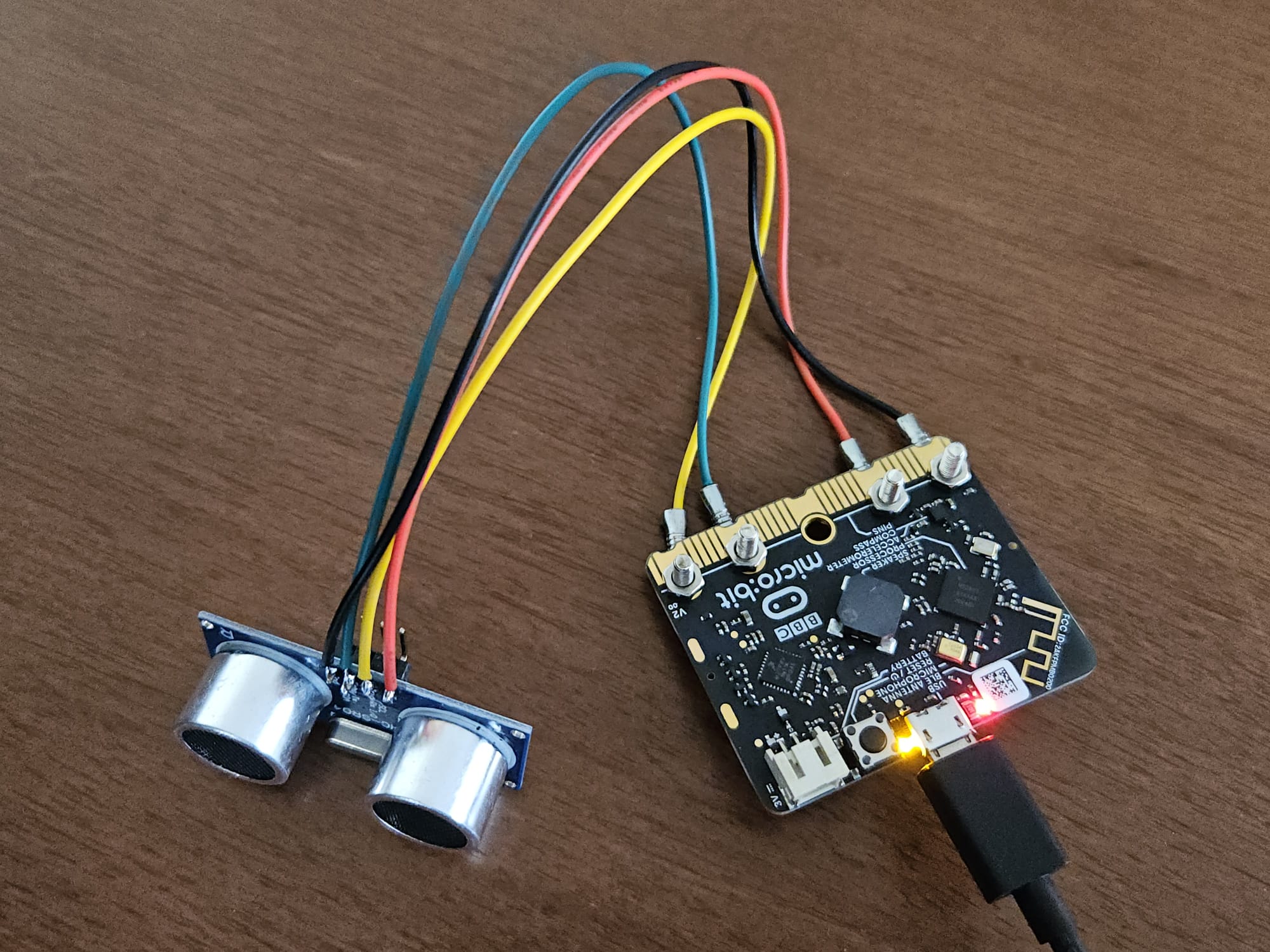}
\includegraphics[width=0.59\textwidth]{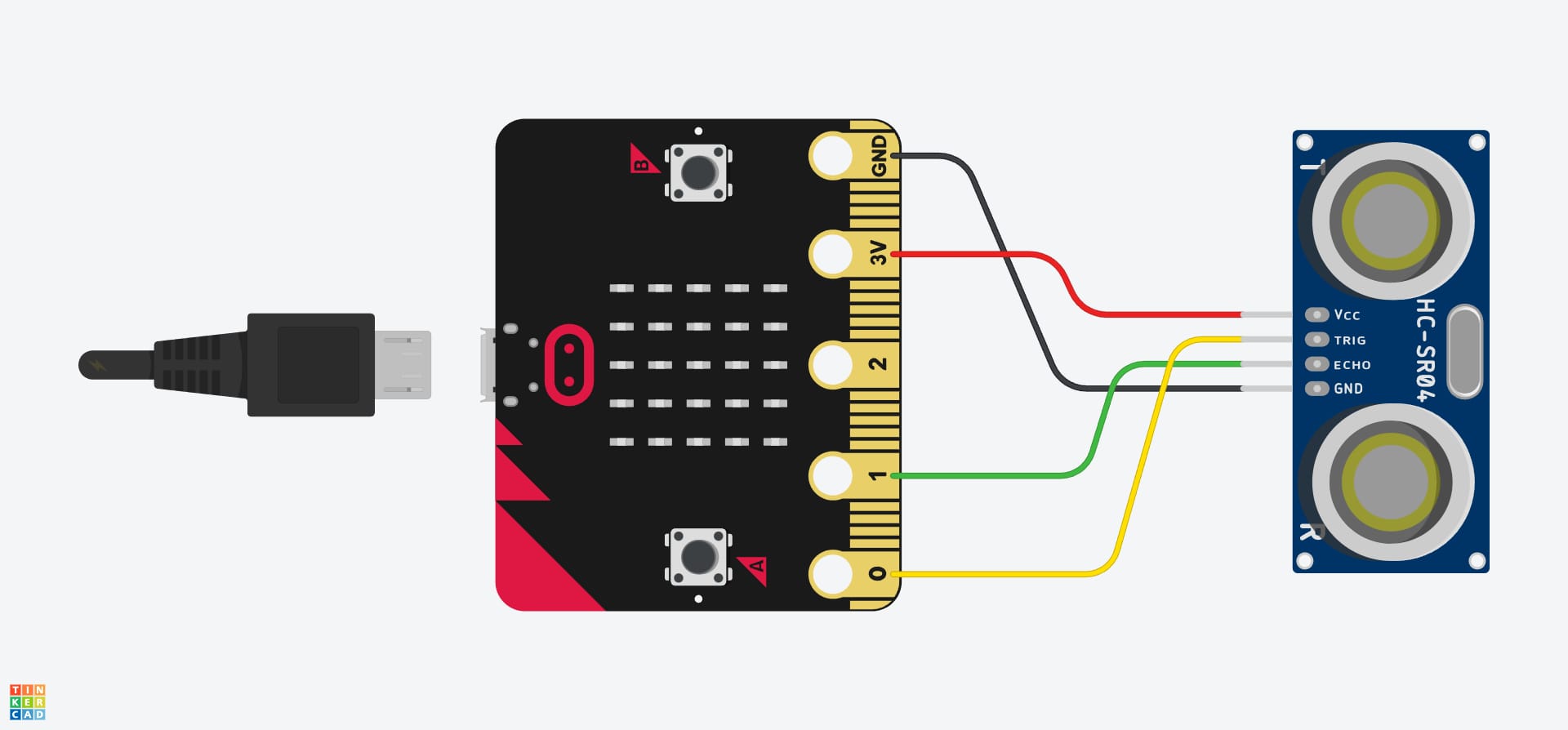}
 \caption{Connection setup used in MissionMotion, showing the micro:bit board connected to an ultrasonic distance sensor and interfaced with a laptop for real-time motion acquisition. When using micro:bit, the HC-SR04P ultrasonic sensor must be employed instead of the HC-SR04, as it is compatible with the 3.3 V operating voltage of the micro:bit.  \label{fig:conexion}}
\end{figure}

\begin{figure}[htbp]
\centering
\includegraphics[width=0.85\textwidth]{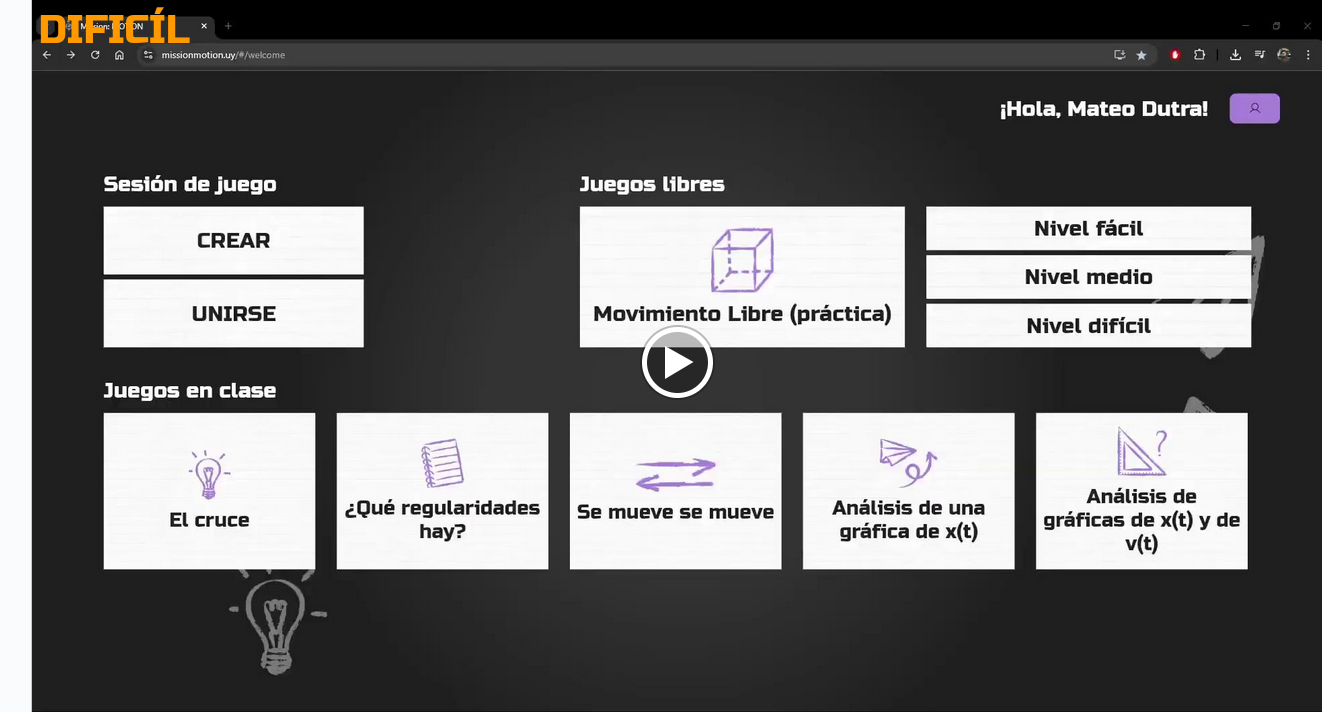}
\caption{Main screen of MissionMotion displaying the activity menu, where students can choose among multiple motion-replication challenges and different input modes. The platform supports both individual gameplay and a classroom-oriented collective mode: teachers have access to a dedicated login that allows them to create virtual classrooms, invite students via code, and view all student results. The teacher menu also provides suggested activities and instructional sequences to support classroom implementation.}
\label{fig:juego2}
\end{figure}

\begin{figure}[htbp]
 \centering
\includegraphics[width=0.48\textwidth]{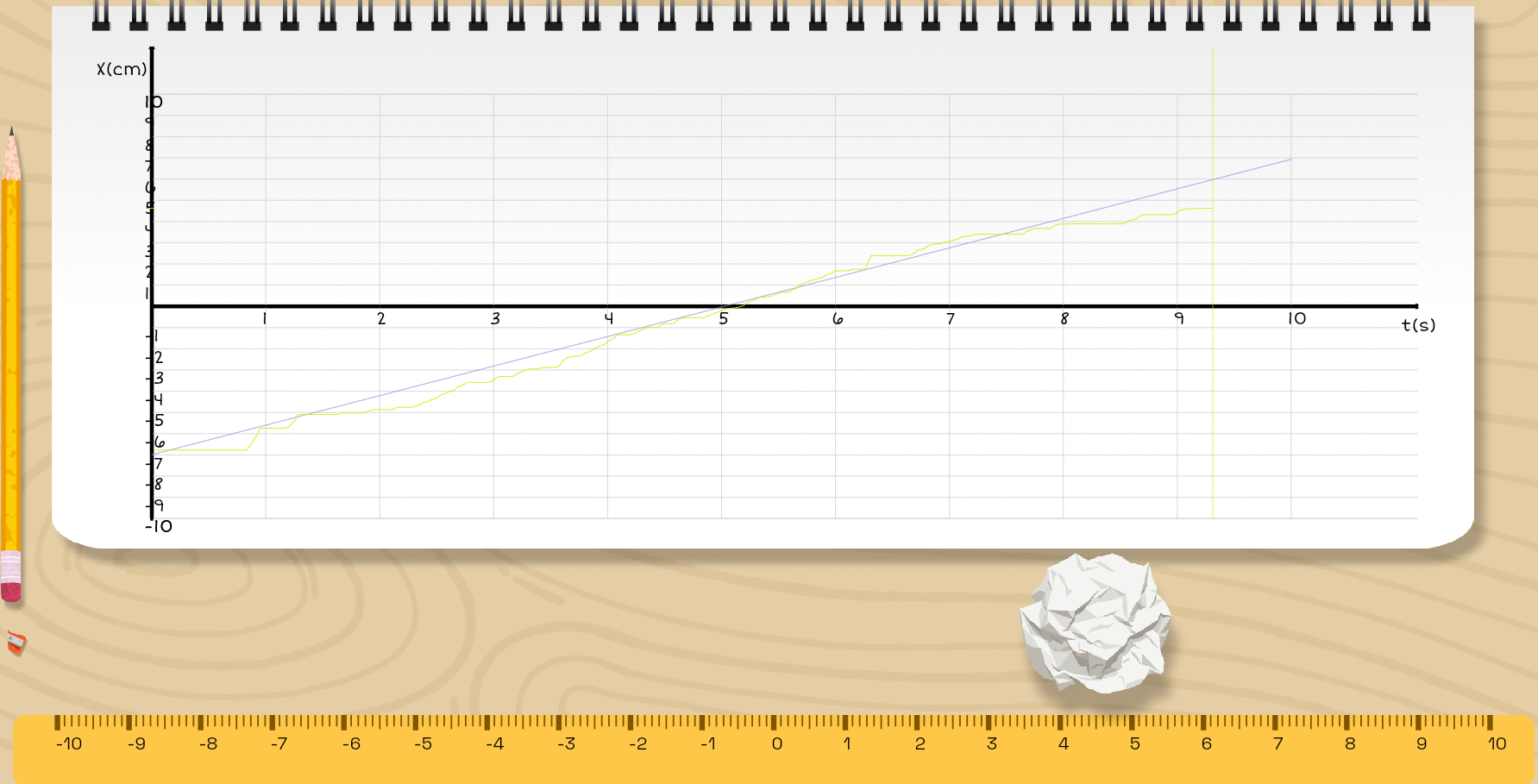}
\includegraphics[width=0.48\textwidth]{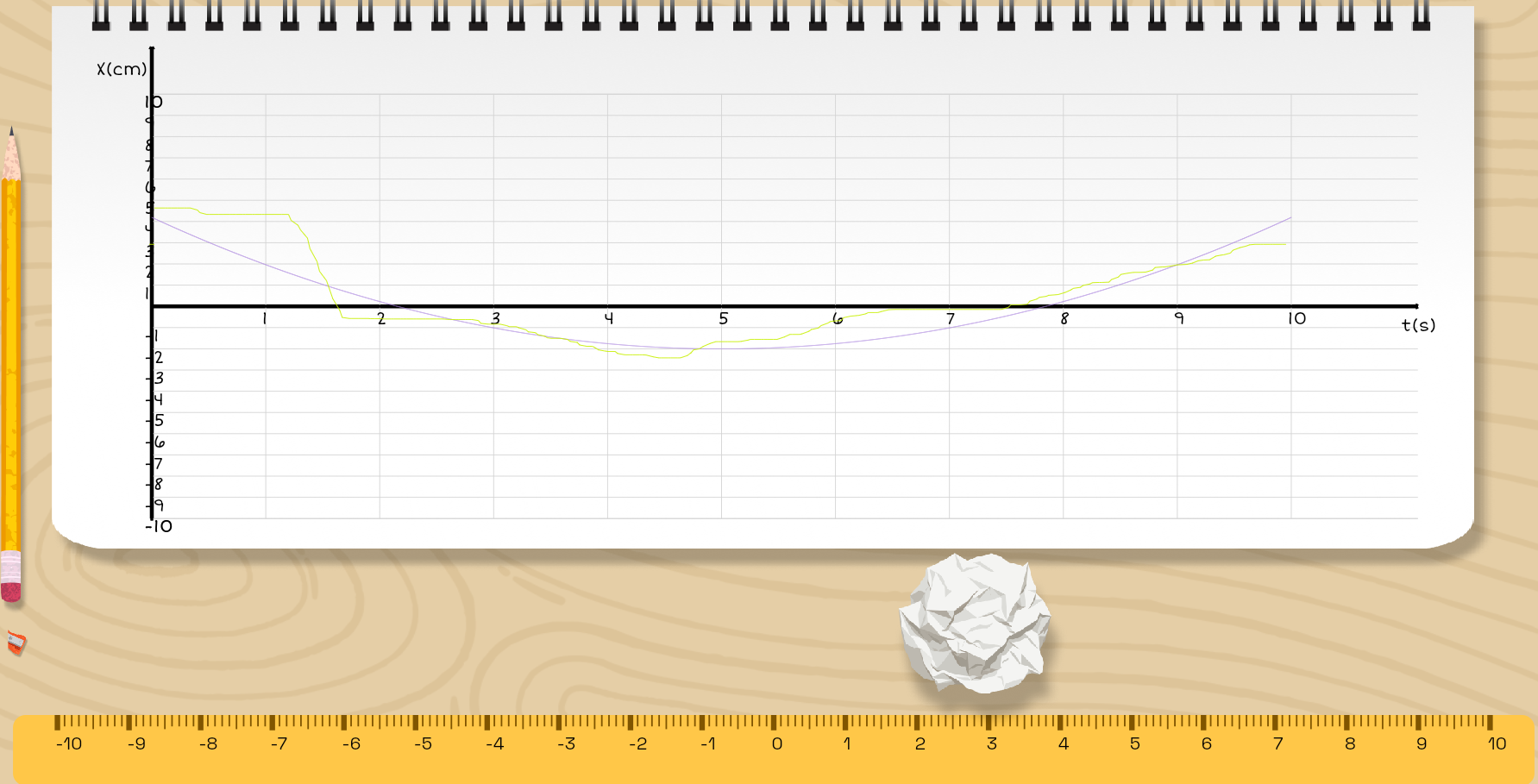}
\includegraphics[width=0.48\textwidth]{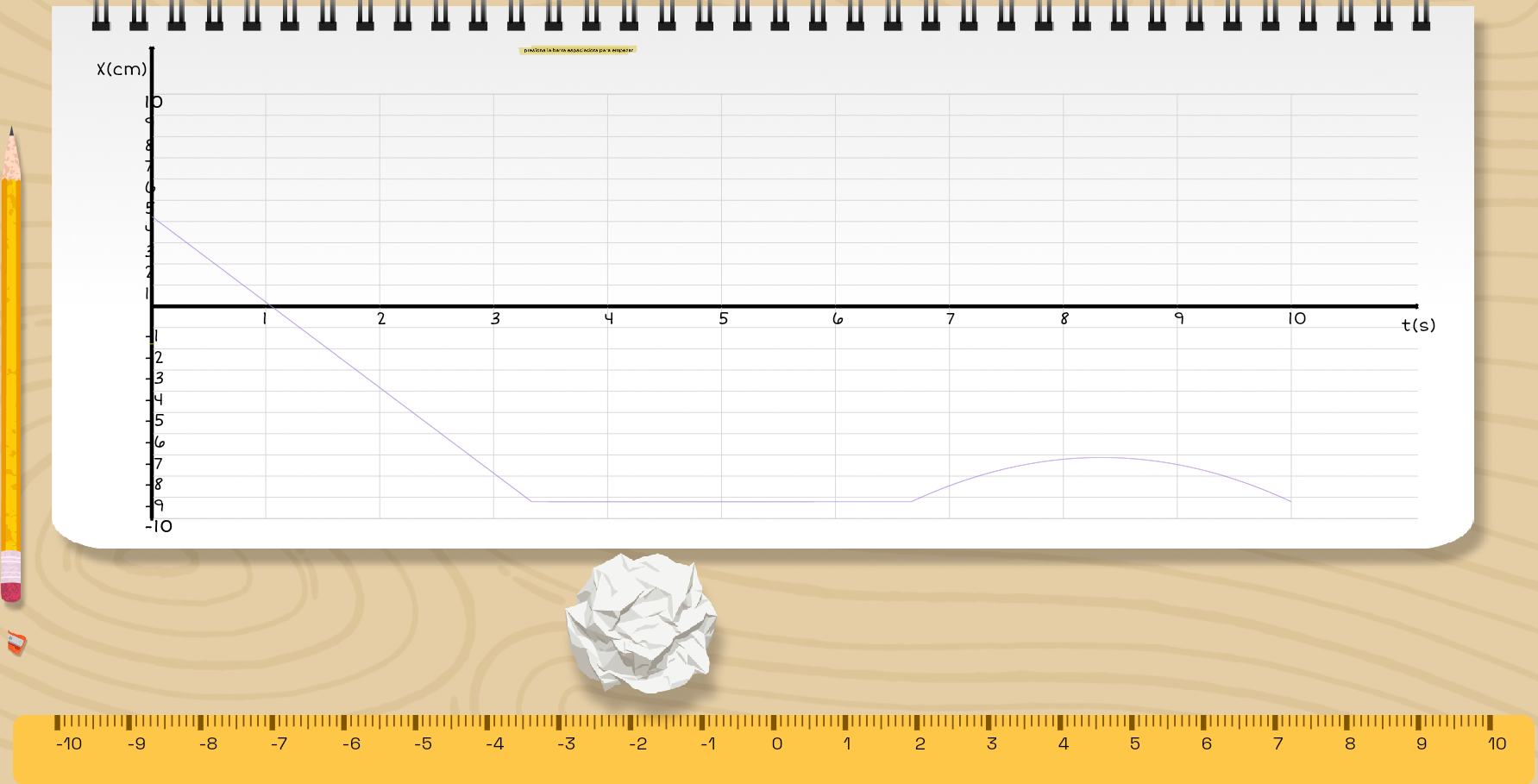}
\includegraphics[width=0.48\textwidth]{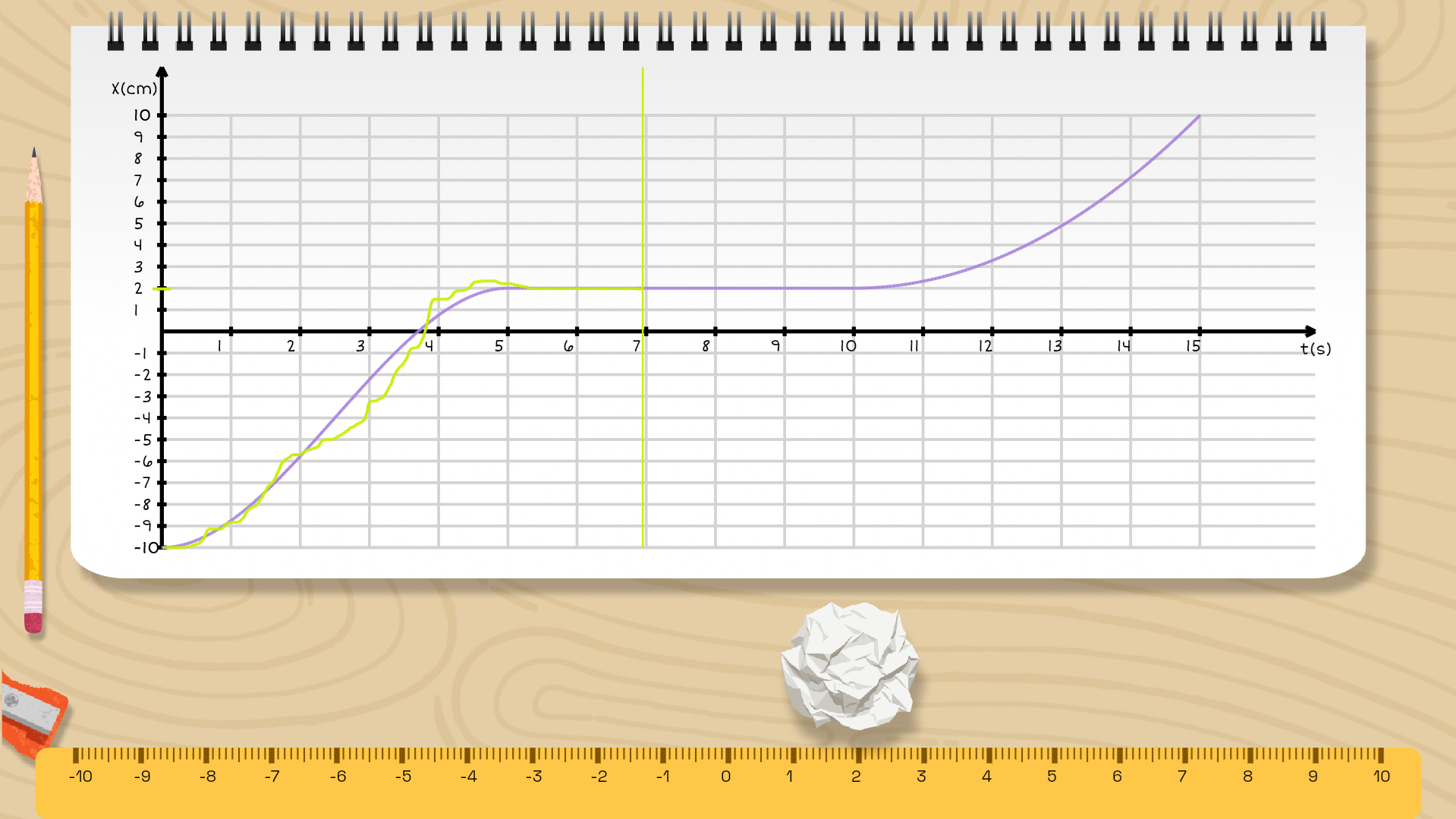}
\caption{Four screenshots of gameplay scenes with different difficulty levels. The player must reproduce the purple position–time graph by moving the mouse or trackpad, represented visually by the paper-ball avatar at the bottom of the screen. As the player moves the mouse, the avatar’s motion is displayed in real time and a yellow graph of the user-generated trajectory is drawn. At the end of the attempt, the system provides a score based on the similarity between the target and the player’s graph.}  \label{fig:juego1}
\end{figure}

To illustrate the versatility of the platform, we offer students a set of representative learning activities implemented within MissionMotion. These tasks invite students to reproduce position–time graphs using their own movement, receiving immediate visual feedback and performance scores based on accuracy. Interaction modes range from simple keyboard, mouse, or touchscreen control to more advanced configurations using ultrasonic sensors or programmed robotic systems, allowing teachers to tailor the experience to available resources and instructional goals. As shown in Figure~\ref{fig:juego2}, the main interface displays the available activity modes, whereas Figure~\ref{fig:juego1} illustrates an example game scenario in which students attempt to match a target motion graph; once the task is completed, the system provides a score based on the similarity between the proposed and player-generated graphs.

The design of MissionMotion is informed by specific, well-documented difficulties in students’ interpretation of kinematics graphs. In particular, the requirement to continuously adjust motion in order to match a target position–time graph makes control of the slope an explicit and unavoidable aspect of the task. Students must anticipate changes in motion to improve their score, implicitly engaging with velocity as a rate of change rather than as a static quantity. The real-time comparison between the target and the user-generated graph makes mismatches immediately visible, supporting iterative refinement across repeated attempts. Rather than treating graphs as static representations to be interpreted after the fact, students interact with them as dynamic objects that respond directly to their actions. From an instructional perspective, this allows teachers to structure activities that emphasize individual exploration or collective discussion, depending on classroom goals and available time. The following section presents results from a pilot classroom implementation, focusing on students’ experience with the environment and on dimensions such as usability, engagement, and perceived educational value, as assessed using the MEEGA+ instrument.

\section{Results of the MEEGA+ test}

In 2025, we conducted a pilot test of the physical–computational environment with five 9th-grade groups from different educational settings: a public high school in Montevideo, a public high school outside the capital, and a private high school in Montevideo. The intervention consisted of a two-hour session in which students interacted with the game using only a mouse to control the character. Activities were structured in two phases: an initial stage of individual exploration, where each participant played at their own pace, followed by a collective phase in which students competed on common levels. This design allowed us to observe both autonomous gameplay and emerging dynamics of competition, problem-solving strategies, and social motivation.

To evaluate this stage, we implemented the MEEGA+ questionnaire (Model for the Evaluation of Educational Games) \cite{petri2016meega}, an internationally validated instrument for analyzing learning experiences mediated by games. MEEGA+ assesses key dimensions of educational gaming environments, including fun and challenge, perceived learning, ease of use, cooperation and competition, as well as immersion and attention levels. We selected this tool because it provides a comprehensive diagnostic framework segmented into specific factors, allowing us to identify concrete opportunities for improvement in the game design and in the classroom dynamics surrounding its use. The insights obtained from this evaluation constitute the basis for further refinement of the environment and for planning new implementation cycles in the upcoming academic year. It is important to note that MEEGA+ does not directly measure conceptual learning gains in kinematics. Instead, it provides insight into affective, motivational, and usability dimensions that are known to condition the effectiveness of active learning environments

\begin{figure}
 \centering
 \includegraphics[width=0.9\textwidth]{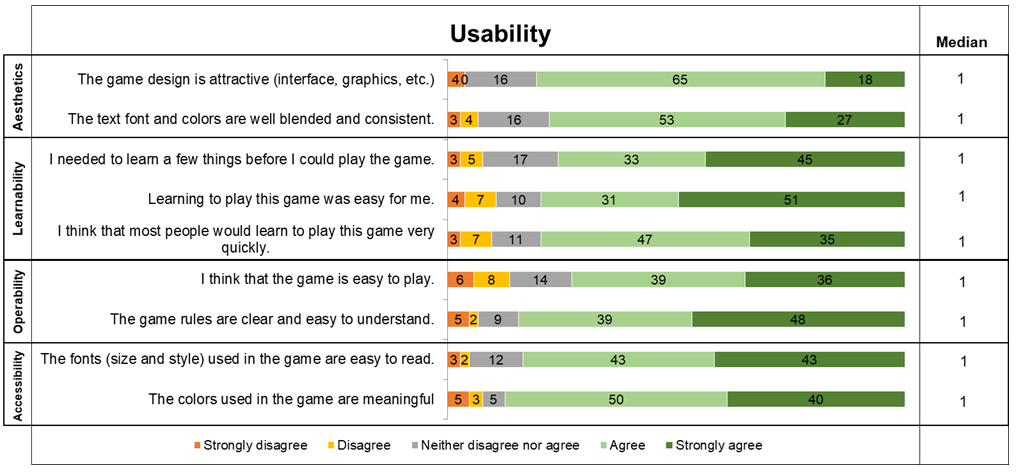}
\caption{Usability evaluation results obtained with the MEEGA+ instrument. The figure summarizes student ratings across dimensions such as ease of use, clarity of interface, perceived control, and feedback effectiveness.  \label{fig:1}}
\end{figure}

\begin{figure}
 \centering
  \includegraphics[width=0.9\textwidth]{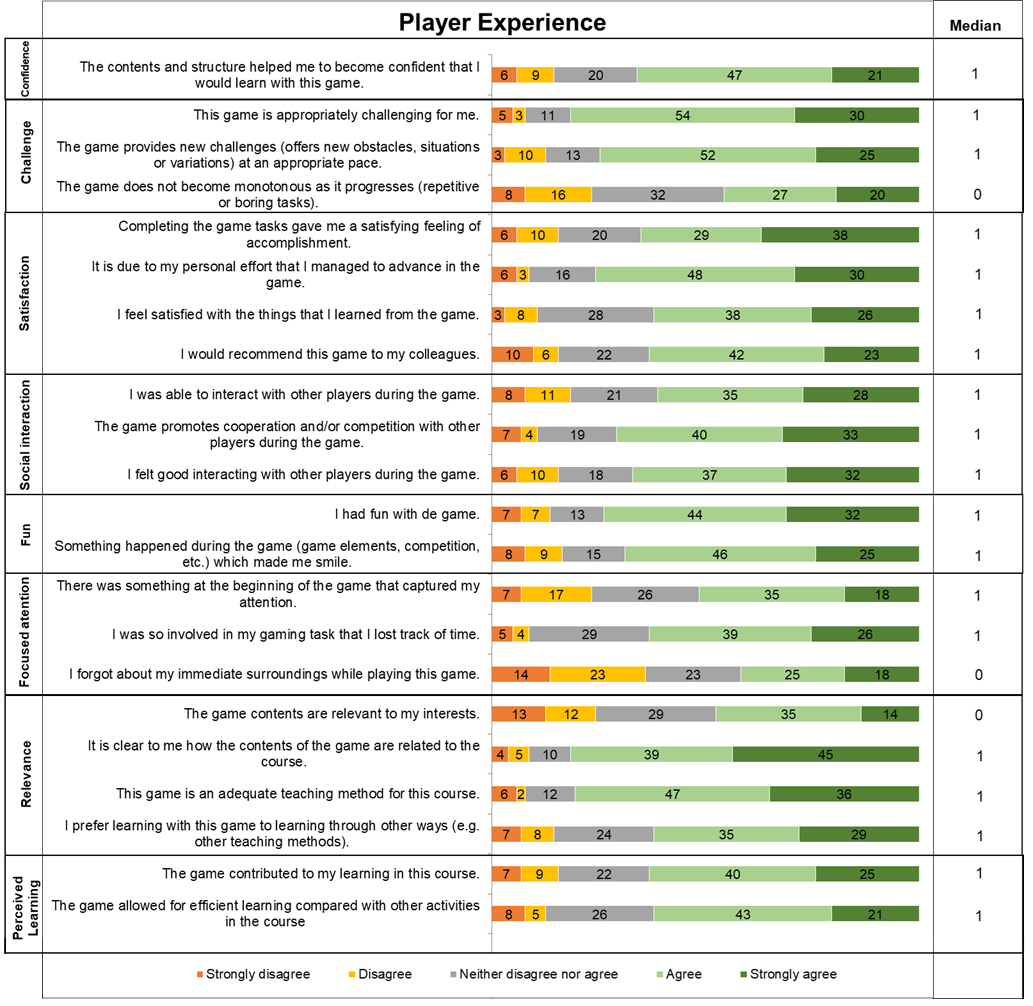}
 \caption{Player experience results from the MEEGA+ assessment. Scores reflect students’ perceptions of enjoyment, challenge, engagement, social interaction, and immersion while interacting with the game environment.  \label{fig:2}}
\end{figure}

The main results of the application of the test are summarized in Figs. \ref{fig:1} and \ref{fig:2}. Open-ended responses revealed that students consistently highlighted the game as fun, visually clear, and motivating. They particularly valued the real-time feedback and the competitive elements, noting that group challenges increased participation and engagement. Classroom observations aligned with these perceptions: students intentionally adjusted their movements to better match the target graph, spontaneous discussions emerged about strategies to “beat the score,” and—remarkably—even students who typically remained passive became actively involved throughout the session.

Teacher feedback further supported these findings. One instructor who implemented the game in his classroom remarked: \begin{quote}
Mission Motion is an excellent tool for classroom work. Not only were my students highly motivated, but reflections about graphs appeared that would rarely surface in a traditional lesson. By trying to reproduce motions, they realized how velocity should change—something that is difficult to notice when only interpreting a graph. The fact that they engaged their own movement promoted deeper reflection about motion characteristics.                                                                                                                                                                                                                                                                                                                                                                                                                           \end{quote}

As shown by the collected data, teachers’ evaluations of the workshop experience were highly positive. In addition to favorable quantitative scores, open-ended responses included enthusiastic comments about the potential of the game for classroom use. Overall, we observed strong motivation among participating teachers, many of whom expressed interest in integrating the tool into their courses next year.

\section{Closing remarks}

In summary, this theoretical framework situates the proposed project at the intersection of physics education research, gamification, and computational thinking. It responds to well-documented learning difficulties in kinematics by leveraging active learning, real-time visualization, and interactive digital technologies. The approach aligns with current educational reforms emphasizing interdisciplinary competencies and seeks to contribute evidence on how playful, technology-rich environments can foster both conceptual and computational understanding in physics. The present pilot study, based on the MEEGA+ evaluation of player experience, constitutes a first step in this direction, providing initial information about how students and teachers interact with the environment in authentic classroom settings.

Grounded in Pierre-Gilles de Gennes’ critique that science education too often begins with abstraction rather than experience, this initiative aims to reintroduce hands-on exploration into physics learning. By bridging embodied experience, game-based learning, and computational thinking, it seeks to help students connect physical concepts to the tangible world, transforming physics from a set of abstract formulas into a lived, interactive experience.

\section{Acknowledgments} We acknowledge financial support from grant
FSED\_2\_2023\_1\_179226 Fundación Ceibal - ANII (Uruguay).

\bibliographystyle{iopart-num}
\bibliography{mybib}

@article{weller2022development,
  title={Development and illustration of a framework for computational
                  thinking practices in introductory physics},
  author={Weller, Daniel P and Bott, Theodore E and Caballero, Marcos
                  D and Irving, Paul W},
  journal={Physical Review Physics Education Research},
  volume={18},
  number={2},
  pages={020106},
  year={2022},
  publisher={APS}
}

@article{Ivanjek2016,
  author    = {Ivanjek, L. and Susac, A. and Planinic, M. and Andrasevic, A. and Milin-Sipus, Z.},
  title     = {Student Reasoning about Graphs in Different Contexts: Kinematics, Mathematics, and Real-World Problems},
  journal   = {Physical Review Physics Education Research},
  volume    = {12},
  number    = {1},
  pages     = {010106},
  year      = {2016},
  doi       = {10.1103/PhysRevPhysEducRes.12.010106}
}

@article{Hake1998,
  author    = {Hake, Richard R.},
  title     = {Interactive-Engagement versus Traditional Methods: A Six-Thousand-Student Survey of Mechanics Test Data for Introductory Physics Courses},
  journal   = {American Journal of Physics},
  volume    = {66},
  number    = {1},
  pages     = {64--74},
  year      = {1998},
  doi       = {10.1119/1.18809}
}

@article{petri2016meega,
  title={MEEGA+: an evolution of a model for the evaluation of educational games},
  author={Petri, Giani and von Wangenheim, C Gresse and Borgatto, Adriano Ferretti},
  journal={INCoD/GQS},
  volume={3},
  pages={1--40},
  year={2016}
}

@article{Duijzer2019,
  author    = {Duijzer, A. C. G. and van den Heuvel-Panhuizen, M. and
                  Veldhuis, M. and Doorman, M. and Leseman, P. P. M.},
  title     = {Supporting Primary School Students’ Reasoning about
                  Graphs through Computer-Based Inquiry Learning with
                  Real-Time Graphs of Their Own Motion},
  journal   = {Journal of Computer Assisted Learning},
  volume    = {35},
  number    = {5},
  pages     = {603--619},
  year      = {2019},
  doi       = {10.1111/jcal.12364}
}

@article{orban2020computational,
  title={Computational thinking in introductory physics},
  author={Orban, CM and Teeling-Smith, RM},
  journal={The Physics Teacher},
  volume={58},
  number={4},
  pages={247--251},
  year={2020},
  publisher={AIP Publishing}
}

@article{dutra2025code,
  title={Code in Motion: Integrating Computational Thinking with Kinematics Exploration},
  author={Dutra, Mateo and Su{\'a}rez, {\'A}lvaro and Marti, Arturo C},
  journal={The Physics Teacher},
  volumen={accepted},
  pages={arXiv:2503.03850},
   year={2025}
}

@article{gambrell2024analyzing,
  title={Analyzing interviews on computational thinking for
                  introductory physics students: Toward a generalized
                  assessment},
  author={Gambrell, Justin and Brewe, Eric},
  journal={Physical Review Physics Education Research},
  volume={20},
  number={1},
  pages={010128},
  year={2024},
  publisher={APS}
}

@article{wing2017computational,
  title={Computational thinking’s influence on research and education for all},
  author={Wing, Jeannette},
  journal={Italian journal of educational technology},
  volume={25},
  number={2},
  pages={7--14},
  year={2017},
  publisher={Edizioni Menab{\`o}-Menab{\`o} srl}
}

@article{basu2016identifying,
  title={Identifying middle school students’ challenges in
                  computational thinking-based science learning},
  author={Basu, Satabdi and Biswas, Gautam and Sengupta, Pratim and
                  Dickes, Amanda and Kinnebrew, John S and Clark,
                  Douglas},
  journal={Research and practice in technology enhanced learning},
  volume={11},
  number={1},
  pages={13},
  year={2016},
  publisher={Springer}
}

@inproceedings{hutchins2018studying,
  title={Studying synergistic learning of physics and computational
                  thinking in a learning by modeling environment},
  author={Hutchins, Nicole and Biswas, Gautam and Conlin, Luke and
                  Emara, Mona and Grover, Shuchi and Basu, Satabdi},
  booktitle={Proceedings of the 26th International Conference on
                  Computers in Education. Philippines: Asia-Pacific
                  Society for Computers in Education},
  year={2018}
}

@article{jara2018policies,
  title={Policies and practices for teaching computer science in Latin America},
  author={Jara, Ignacio and Hepp, Pedro and Rodriguez, Jaime},
  journal={Microsoft},
  year={2018}
}

@article{chevalier2020fostering,
  title={Fostering computational thinking through educational
                  robotics: A model for creative computational problem
                  solving},
  author={Chevalier, Morgane and Giang, Christian and Piatti, Alberto
                  and Mondada, Francesco},
  journal={International journal of STEM education},
  volume={7},
  number={1},
  pages={39},
  year={2020},
  publisher={Springer}
}

@article{caballero2018prevalence,
  title={Prevalence and nature of computational instruction in
                  undergraduate physics programs across the United
                  States},
  author={Caballero, Marcos D and Merner, Laura},
  journal={Physical Review Physics Education Research},
  volume={14},
  number={2},
  pages={020129},
  year={2018},
  publisher={APS}
}

@article{mcdermott1987student,
  title={Student difficulties in connecting graphs and physics:
                  Examples from kinematics},
  author={McDermott, Lillian C and Rosenquist, Mark L and Van Zee, Emily H},
  journal={American journal of physics},
  volume={55},
  number={6},
  pages={503--513},
  year={1987},
  publisher={American Association of Physics Teachers}
}

@article{trowbridge1980investigation,
  title={Investigation of student understanding of the concept of
                  velocity in one dimension},
  author={Trowbridge, David E and McDermott, Lillian C},
  journal={American journal of Physics},
  volume={48},
  number={12},
  pages={1020--1028},
  year={1980},
  publisher={American Association of Physics Teachers}
}

@article{aprendizajeactivoB,
  title={Peer instruction: Ten years of experience and results},
  author={Crouch, Catherine H and Mazur, Eric},
  journal={American journal of physics},
  volume={69},
  number={9},
  pages={970--977},
  year={2001},
  publisher={American Association of Physics Teachers}
}

@article{aprendizajeactivoC,
  title={RealTime Physics: active learning labs transforming the
                  introductory laboratory},
  author={Sokoloff, David R and Laws, Priscilla W and Thornton, Ronald K},
  journal={European journal of physics},
  volume={28},
  number={3},
  pages={S83},
  year={2007},
  publisher={IOP Publishing}
}

@article{aprendizajeactivoD,
  title={Learning motion concepts using real-time microcomputer-based
                  laboratory tools},
  author={Thornton, Ronald K and Sokoloff, David R},
  journal={American journal of Physics},
  volume={58},
  number={9},
  pages={858--867},
  year={1990},
  publisher={AIP Publishing}
}

@article{trowbridge1981investigation,
  title={Investigation of student understanding of the concept of
                  acceleration},
  author={Trowbridge, David E and McDermott, Lillian C},
  journal={American journal of Physics},
  volume={49},
  number={3},
  pages={242--253},
  year={1981}
}

@article{beichner1994testing,
  title={Testing student interpretation of kinematics graphs},
  author={Beichner, Robert J},
  journal={American journal of Physics},
  volume={62},
  number={8},
  pages={750--762},
  year={1994},
  publisher={[Woodbury, NY, etc. Published for the American Association of Physics~…}
}

\end{document}